\documentclass[aps,prl,twocolumn,groupedaddress,showkeys]{revtex4}
\usepackage{graphics}

\begin{document}

\title{Network information and connected correlations}

\author{Elad Schneidman,$^{1-3}$ Susanne Still,$^{1,3}$ Michael J. Berry II,$^{2}$ and William Bialek$^{1,3}$}
\affiliation{Departments of $^{1}$Physics and $^{2}$Molecular Biology, and\\
 $^{3}$Lewis--Sigler Institute for Integrative Genomics\\
 Princeton University, Princeton, New Jersey
08544 USA}

\date{\today}

\begin{abstract}
Entropy and information provide natural measures of correlation among elements in a
network.  We construct here the information theoretic analog of connected correlation functions:
irreducible $N$--point correlation is measured by a decrease in entropy for the joint
distribution of $N$ variables relative to the maximum entropy allowed by all the observed
$N-1$ variable distributions. We calculate the ``connected information'' terms for several examples, and show that it also enables the decomposition of the information that is carried by a population of elements about an outside source.  
\end{abstract}

\maketitle

In statistical physics and field theory, the nature of order in a
system is characterized by correlation functions.  These ideas are
especially powerful because there is a direct relation between the
correlation functions and experimental observables such as
scattering cross sections and susceptibilities. As we move toward
the analysis of more complex systems, such as the interactions
among genes or neurons in a network, it is not obvious how to
construct correlation functions which capture the underlying
order.  On the other hand it is possible to observe directly  the
activity of many single neurons in a network or the expression
levels of many genes, and hence real experiments in these systems
are more like Monte Carlo simulations, sampling the distribution
of network states.

Shannon proved that, given a probability distribution over a set of variables,
entropy is the unique measure of what can be learned by observing these variables,
given certain simple and plausible criteria (continuity, monotonicity and additivity) \cite{Shannon-48}.
By the same arguments, mutual information arises as the unique measure of the interdependence of two variables, or two sets of variables. Defining information theoretic analogs of higher order correlations has proved to be more difficult \cite{McGill-54,Watanabe-60,Good-63,Martignon-00,Amari-01,Csiszar-75,Han-78,sv-98,Bell-02}. When we compute $N$--point correlation functions in statistical physics
and field theory, we are careful to isolate the connected
correlations, which are the components of the $N$--point
correlation that cannot be factored into correlations among groups
of fewer than $N$ observables.
We propose here an analogous measure of ``connected information'' which generalizes
precisely our intuition about connectedness and interactions from
field theory; a closely related discussion for quantum information
has been given recently \cite{lpw02}.

Consider $N$ variables $\{x_{\rm i}\}, i=1,2,...,N$, drawn from the
joint probability distribution $P(\{x_{\rm i}\})$; this has an entropy \cite{sums}.
\begin{equation}
S(\{x_{\rm i}\}) = -\sum_{\{x_{\rm i}\}} P(\{x_{\rm i}\}) \log P(\{x_{\rm i}\}).
\end{equation}
The fact that $N$ variables are correlated
means that the entropy $S(\{x_{\rm i}\})$ is smaller than the sum of
the entropies for each variable individually,
\begin{equation}
S(\{x_{\rm i}\})  <  \sum_{\rm i} S(x_{\rm i}).
\end{equation}
The total difference in entropy between the interacting
variables and the variables taken independently can be written as \cite{McGill-54,Watanabe-60}
\begin{eqnarray}
I(\{x_{\rm i}\}) &\equiv& \sum_{\rm i} S(x_{\rm i}) - S(\{x_{\rm i}\}) \nonumber \\
&=& \sum_{\{x_{\rm i}\}} P(\{x_{\rm i}\}) \log\left[
{P(\{x_{\rm i}\})}\over{\prod_{\rm j}P_{\rm j}(x_{\rm j})}\right] ,
\label{KL1}
\end{eqnarray}
which  is the Kullback--Leibler divergence between the true distribution $P(\{ x_{\rm
i}\})$ and the ``independent'' model formed by taking the product of the marginals, $\prod_{\rm j}P_{\rm j}(x_{\rm j})$. This has been
called the {\em multi--information};  it provides a general measure of non--independence
among multiple variables in a network.

The multi--information alone does not tell us how much of the
non--independence among $N$ variables is intrinsic to the full $N$ variables and how much can be
explained from pairwise, triple, and higher order interactions.  For example, if the $x_{\rm i}$'s are binary
variables or equivalently Ising spins $\sigma_{\rm i}$, and if the full distribution
$P(\{ \sigma_{\rm i} \})$ is a conventional Ising model with pairwise exchange
interactions, then in an obvious sense there is nothing ``new'' to learn by observing
triplets of spins that can't be learned by looking at all the pairs.  On the other hand,
if $\sigma_3$ is formed as the exclusive OR (XOR) of the variables $\sigma_1$ and $\sigma_2$,
then  the essential structure of $P(\sigma_1, \sigma_2, \sigma_3)$ is
contained in a three--spin interaction;  if $\sigma_1$ an $\sigma_2$ are chosen
at random as inputs to the XOR, then all pairwise mutual informations among the $\sigma_{\rm i}$ will be zero, although the multi--information will be one bit (Fig. 1)

What we would like to do in our example of three variables is to separate that component of $I(x_{1};x_{2};x_{3})$ which is expected from observations on pairs of variables from that component which is intrinsic to the triplet. Observing the variables in pairs means that we can construct all of the pairwise marginals
$P_{\rm ij} = \sum_{x_{\rm k}} P(x_{\rm i}, x_{\rm j}, x_{\rm k})$.
Knowledge of these marginals provides (in general) a partial characterization of 
the full probability distribution $P(x_1, x_2, x_3)$.  Following Jaynes \cite{Jaynes-57a} we can quantify this knowledge by saying that the pairwise marginals set a maximum value of the entropy for the full distribution.  More generally, if we have $N$ variables and we observe all the subsets of $k$ elements, then there is a maximum entropy for the distribution $P(\{x_{\rm i}\})$ that is consistent with all of the $k$--th order marginals.
Let us write  this maximum entropy distribution by $\tilde P^{(k)}(\{x_{\rm i}\})$ and denote the entropy of a probability distribution by $S[P]$; note that
\begin{equation}
\tilde P^{(1)}(\{x_{\rm i}\}) =   \prod_{{\rm i}=1}^{n}P_{\rm i}(x_{\rm i}),
\end{equation}
and that $\tilde P^{(N)}(\{x_{\rm i}\})$ is just the true distribution $P(\{x_{\rm i}\})$.
Then we can decompose the multi--information among the $N$ variables  into a sequence of terms:
\begin{eqnarray}
\label{Eq:Multi-information-Break}
I(\{x_{\rm i}\})  & \equiv &
 S\left[\prod_{i=1}^{N}P_{i}(x_{i})\right]  - S[P(\{x_{\rm i}\})]  \nonumber \\
 & =  & \sum_{k=2}^{N}I_{C}^{(k)}(\{x_{\rm i}\}) ,
\end{eqnarray}
where we define the {\em connected information of order $k$,}
\begin{equation}
I_{C}^{(k)}(\{x_{\rm i}\})  =  S[\tilde{P}^{(k-1)}(\{x_{\rm i}\})] -S[\tilde{P}^{(k)}(\{x_{\rm i}\})]  .
\end{equation}
The connected information of order $k$ is positive or zero; it represents the amount by which the maximum possible entropy of the system decreases when we go from knowing only the marginals of order $k-1$ to knowing also the marginals of order $k$.  Each time that we increase the number of elements that we can observe simultaneously we uncover a potentially  richer set of correlations, leading to a reduction in the maximum possible entropy; the connected information measures this entropy reduction.

Computing the connected information requires that we construct the maximum entropy distributions consistent with marginals of order $k$.  In general this is a difficult problem.  Recall that to maximize the entropy when we know the expectation values of functions $F_\mu (\{x_{\rm i}\})$, the resulting probability distribution is of the Boltzmann form, $P(\{ x_{\rm i}\}) \propto \exp[-\sum_\mu \lambda_\mu F_\mu (\{x_{\rm i}\})]$, where the $\lambda_\mu$ are Lagrange multipliers conjugate to each function \cite{Jaynes-57a}. 
We can think of each marginal distribution as a set of expectation values over the full distribution, so that we need one Lagrange multiplier for each  k--tuple of $x$ values. The distribution $\tilde{P}^{(k)}$ thus has the form of a Boltzmann distribution with $k$--body interactions; these interactions are arbitrary functions which have to be determined by matching the observed marginals. As an example, for three variables with known pairwise marginals the maximum entropy distribution takes the form
\begin{eqnarray}
\tilde{P}^{(2)}(x_1, x_2, x_3)  &=& {1\over Z} \exp [ - \lambda_{12}(x_1,x_2) \nonumber\\
&& - \lambda_{23}(x_2,x_3) - \lambda_{31}(x_3,x_1) ] .
\end{eqnarray}
\noindent For a physical system that has at most $K$--body interactions among the $N$ variables, $P^{(K)}$ will be the exact distribution.  Correspondingly, $I_{C}^{(k)} = 0$ for $k > K$.

In general the functions $\lambda$ are difficult to determine from the observed marginals, but this is not the case for $k=1$. This is a well known but important point:  the maximum entropy distribution consistent with one--body marginals is just the product of the marginals, but the maximum entropy distribution consistent even with two--body (pairwise) marginals is {\em not} simply written in terms of the marginals because the observed two--body correlations include an average over interactions with all other degrees of freedom.  As a result, even the second order maximum entropy distributions for $N$ variables are not simply related to the pairwise marginals, and  the second order connected information is not simply related to the mutual information among pairs of variables; $I_C^{(2)}$  is larger than the mutual information between any pair of variables, but is {\em not} equal  to their sum. 

The fact that maximum entropy distributions have an exponential form, and in the binary or Ising case this form includes only a finite set of parameters, connects our discussion with previous work.  A number of authors have used the maximum entropy distribution for families of parameterized models as part of statistical tests for  the existence of higher order interactions  \cite{Good-63,Soofi-92,Martignon-00}
In related work, Amari \cite{Amari-01} has constructed a geometry on the parameter space for exponential families using the Fisher information as a metric, and in this geometry the maximum entropy distributions are orthogonal projections onto subspaces of the full parametric space (see also \cite{Csiszar-75}). Rather than providing a parametric model of
$k$--th order interactions and determining a confidence level, the set of $I_{{C}}^{(k)}$ provides a quantitative characterization of the relative importance of various order interactions, independent of parameterization.

\begin{figure}
\vspace{-.25cm}
\centerline{
\resizebox {0.45\textwidth}{!}{\includegraphics{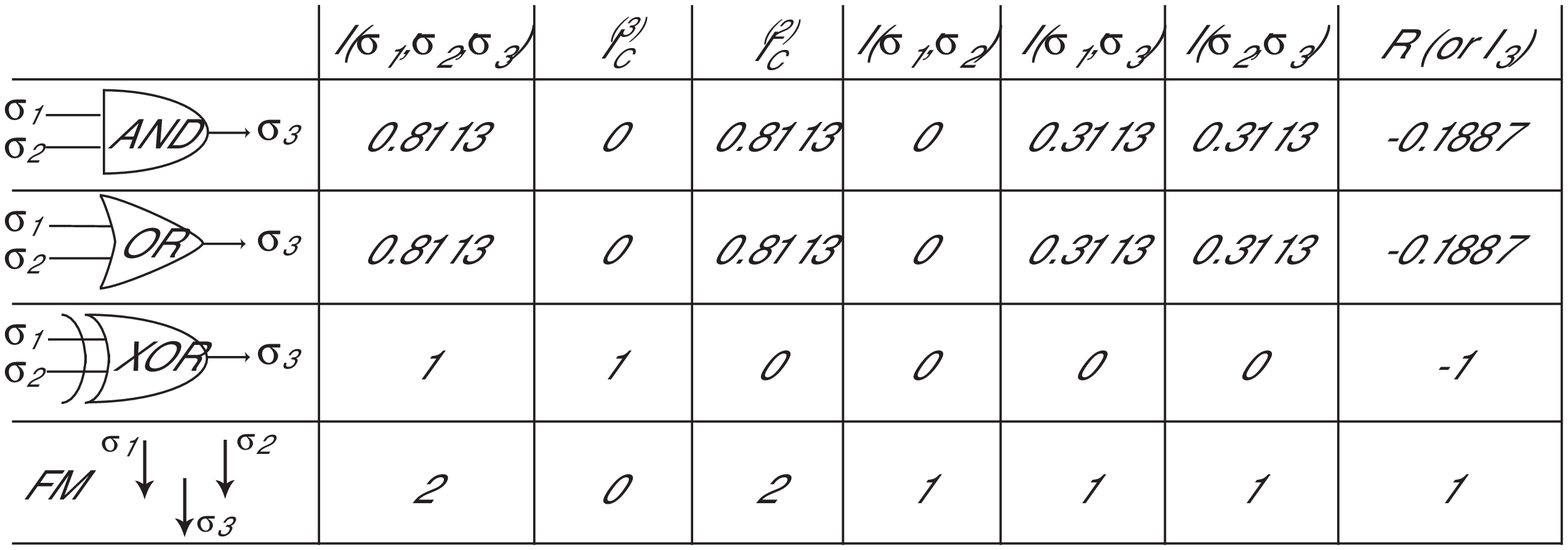}}}
\caption[]{\footnotesize
The values of multi--information, connected--information of orders 2 and 3, the pairwise mutual information and pairwise redundancy for 3 binary variables, whose probability distribution is given by the logical functions AND, OR and XOR (with the inputs $\sigma_{1}$ and $\sigma_{2}$ chosen at random), and the case of ferromagnetic interaction, FM.}
\label{Fig:Multi-information-Table}
\end{figure}

\begin{figure}
\vspace{-.25cm}
\centerline{
\resizebox {0.45\textwidth}{!}{\includegraphics{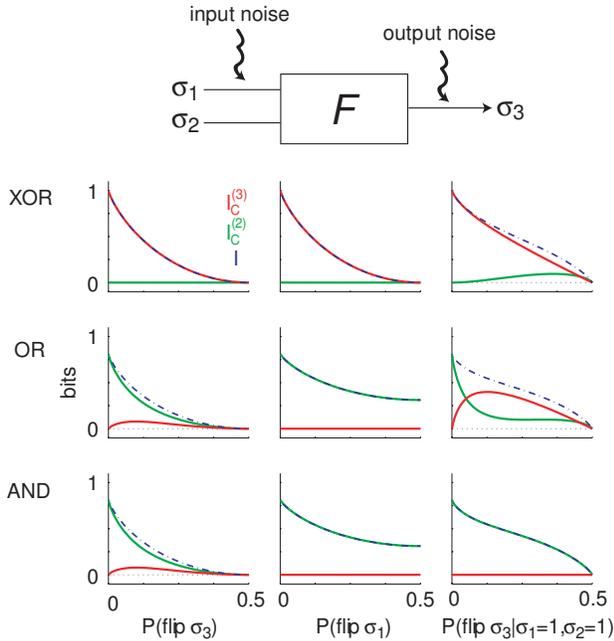}}}
\caption[]{\footnotesize
Correlated--information of orders 2 and 3 and the multi--information for 3 variables whose joint probability distribution is given by noisy logical functions.
Each panel presents the $I_{C}$'s and I values for a noisy version of one boolean gate (XOR in first row, OR in second, AND in third), as a function of noise amplitude. The three types of noise are output noise (probability of flipping $\sigma_{3}$), input noise (probability of flipping $\sigma_{1}$) and input-dependent output noise (probability of flipping $\sigma_{3}$, given that $\sigma_{1}=1$ and $\sigma_{2}=1$). 
}
\label{Fig:Noisy-Logicals}
\end{figure}

As examples (Fig.~\ref{Fig:Multi-information-Table}), consider three binary or Ising variables related either by boolean functions (AND, OR, XOR) or coupled through a pairwise ferromagnetic interactions (FM). For these simple functions, we find that the multi--information is composed of either pure 2-body interactions or pure 3-body ones, as our intuition suggests. When we add noise either to the input or output of the boolean functions (Fig.~\ref{Fig:Noisy-Logicals}) we degrade the correlations, but more interestingly we find that pure 2-body interactions such as AND and OR show a 3-body interaction component for some types of noise (even for noise sources which are state dependent). For the pure 3-body XOR, noise may result in the appearance of 2-body interactions. For these three functions, input noise only changes the strength of the existing interactions, rather than introducing a new kind of effective interaction.

\begin{figure}
\centerline{
\resizebox {0.45\textwidth}{!}{\includegraphics{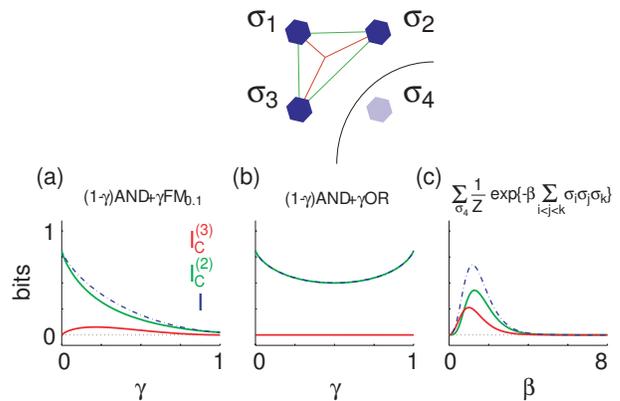}}}
\caption[]{\footnotesize
Correlated--information of orders 2 and 3  and the multi--information, for networks of three binary observable elements, $\sigma_{1},\sigma_{2},\sigma_{3}$, with one hidden binary element $\sigma_{4}$. (a) $I_{C}$'s and I values for  a network where the value of $\sigma_{4}$ determines the pairwise interaction between the other elements: if $\sigma_{4}=0$ then $\sigma_{3}=AND(\sigma_{1},\sigma_{2})$; if $\sigma_{4}=1$ then the interaction among the observable variables is (pairwise) ferromagnetic with a finite temperature ($\beta=0.1$). Information values are plotted as a function of $\gamma = P(\sigma_{4}=0)$.
 $P(\sigma_{1},\sigma_{2},\sigma_{3})$ in
 (b) same as a, but for a $\sigma_{4}$-dependent mixture of AND and OR. For this case there is no effective 3--body interaction. (c) $I_{C}$'s and I for the three observable binary variables network, where the full 4 element network has pure 3-body interactions, plotted as a function of the inverse temperature $\beta$.
 }
\label{Fig:CoInfo-hidden}
\end{figure}

As is familiar in physical examples, if we observe only some of the elements of a network then the effect of the hidden elements may be to create new effective interactions among the observed elements.
As examples (Fig.~\ref{Fig:CoInfo-hidden}), when one hidden binary element determines the nature of pure pairwise interaction among the remaining elements, the observable subnetwork can have an effective 3--body interaction.  
Alternatively, for a network with only pure 3--body interactions, hidden elements can induce an effective 2--body interaction among the observables.

As noted above, the connected information at second order (for example) cannot be written simply in terms of the mutual information among pairs of variables.  Many previous authors have looked for linear combinations of mutual information measures which might provide  measures of higher order interaction, and among these one approach  is of particular interest \cite{McGill-54,Watanabe-60,Han-78,sv-98,Bell-02}:   If we draw a Venn diagram of regions in the plane 
corresponding to the variables $x_1, x_2, x_3$, identifying areas
with entropies, then the mutual information
$I(x_{\rm i}; x_{\rm j})$ between two variables is the area of
their intersection, and  there is a unique region shared by all three
variables; with the area--entropy correspondence the size of
this ``triplet information'' is
\begin{eqnarray} I_3 &=&\sum_{\rm i} S(x_{\rm i}) -
\sum_{\rm i < j} S(x_{\rm i} ,
x_{\rm j}) + S(x_1, x_2, x_3) \nonumber \\
&=& I(x_1; x_2; x_3) - \sum_{\rm i < j} I(x_{\rm i};
x_{\rm j}) .
\end{eqnarray}
This proposal for measuring a pure triplet information has natural
generalizations to more than three variables.

There are at least two difficulties with the triplet information defined by $I_3$ (see a thorough discussion in \cite{Watanabe-60}).  First,
despite the identification of shared information with areas in the plane, we find that $I_{3}$
can be negative (AND, OR and XOR in Fig.~\ref{Fig:Multi-information-Table}). Second, $I_3$
can be nonzero even for networks that have only pairwise interactions (FM in  Fig.~\ref{Fig:Multi-information-Table}).

Rather than ``triplet information,''  $I_{3}$ actually measures   \cite{McGill-54,Watanabe-60}
the information that $x_1$ and $x_2$ together provide about $x_3$ with the information that these two variables provide separately:
\begin{eqnarray}
I_{3} = \left[I(x_1;x_3) + I(x_2;x_3) \right] - I(\{x_1, x_2\};x_{3}).
\label{Eq:PairwiseSynergy}
\end{eqnarray}
This comparative measure of information is symmetric under permutation of the indices, so the labeling of variables as $1,2,3$ is arbitrary.  If $I_{3}$ is positive, then any pair $x_{\rm i}$ and $x_{\rm j}$ are redundant in terms of the information that they provide about the remaining $x_{\rm k}$. If $I_{3}$ is negative, then there is synergy---two variables taken together are more informative than they are when taken separately.

The question of synergy and redundancy brings us back to one of the primary motivations for this analysis.  Consider the responses $x_1, x_2, \cdots , x_N$ of a collection of elements to some stimulus $y$ -- for example a group of neurons responding to a sensory stimulus.  For each neuron $\rm i$  we can ask how much information the response provides about the sensory world, $I(x_{\rm i} ; y)$. When we look at a pair of neurons, we can ask whether these neurons provide redundant or synergistic information (using eq. \ref{Eq:PairwiseSynergy}; see e.g. \cite{gr-93,brenner00}).
Similarly for a large population of neurons we can compare the information in the population, $I(\{x_{i}\}; y)$, with the sum of informations provided by the neurons individually, $\sum_{\rm i} I(x_{\rm i};y)$. This comparison, however, does not tell us whether (for example) the synergy in the population is the result of pairwise correlations or whether there are special combinations of responses across all three or more neurons which provide extra information. The possible significance of such multi--neuron combinatorial events has been discussed for many years (see e.g. \cite{pag88,Abeles-91,schnitzer+meister03}).

We recall that the information provided by a population of neurons can be written as
\begin{equation}
I(\{x_{\rm i}\};y) =  S[P(\{x_{\rm i}\})] - \langle S[P(\{x_{\rm i}\}|y) ]\rangle_{y},
\end{equation}
where $\langle\cdots\rangle_{y}$ denotes an average over the distribution of sensory inputs.
The redundancy of the population is defined as
\begin{equation}
R(\{x_{\rm i}\})  \equiv \sum_{\rm i =1}^N I(x_{\rm i} ; y) - I(\{x_{\rm i}\};y),
\end{equation}
where negative $R$ corresponds to synergy.  We note that $R$ can be written as the difference between two multi--information terms,
\begin{eqnarray}
R(\{x_{\rm i}\}) &=& \left(\sum_{\rm i =1}^N S[P(x_{\rm i})] - S[P(\{x_{\rm i}\})]\right) \nonumber\\
&&-\bigg\langle  \sum_{\rm i =1}^N S[P(x_{\rm i}| y)] - S[P(\{x_{\rm i}\}|y)]\bigg\rangle_y .
\end{eqnarray}
The first term is the multi--information in the distribution of neural responses, which measures the extent to which the total ``vocabulary'' of the population is reduced through correlations, while the second term is the multi--information in the distribution of responses to a given stimulus.  Each of these terms in turn can be expanded as a sum of connected informations, so that
\begin{eqnarray}
R(\{x_{\rm i}\}) = \sum_{k=2}^N \left[ I_{C}^{(k)}(\{x_{\rm i}\}) - \langle I_{C}^{(k)}(\{x_{\rm i}\}|y)\rangle_y\right].
\end{eqnarray}

\noindent where $I_{C}^{(k)}(\{x_{\rm i}\}|y)$ is the connected--information of order $k$ in the network of $\{x_{\rm i}\}$, for a given value of $y$. By analogy with the discussion of synergy in pairs \cite{brenner00}, the terms $I_{C}^{(k)}(\{x_{\rm i}\})$ quantify the contribution of kth order interactions to restricting the vocabulary of the population response (much as not all k letter combinations form words in English), while the terms 
$\langle I_{C}^{(k)}(\{x_{\rm i}\}|y)\rangle_y$ quantify the contribution of kth order correlations to reducing the noise in the population response.

To summarize, the maximum entropy construction of connected information presented here provides us both with a method for decomposing the correlations within a network and for quantifying the contribution of these correlations to the information that network states can provide about external signals.  Since any part of a network can be thought of as `external' to its compliment, this unified discussion of internal correlations and the representation of external signals is attractive.  

This work was supported in part by a Pew Scholar Award and a grant from the E. Mathilda Ziegler Foundation (to MJB),  a Rothschild Foundation/Yad Hanadiv fellowship (to ES), and  by the German Research Foundation (DFG), grant no. Sti197 (to SS).


\begin{thebibliography}{99}
\bibitem{sums}
We write a sum  to indicate sums or integrals
as appropriate.
\end{thebibliography}
\end{document}